\documentclass[jcp,amsmath,amssymb,showpacs,nofootinbib,preprint]{revtex4-1}
\usepackage{amssymb}
\usepackage{amsmath}

\newcommand{\beq}{\begin{equation}}
\newcommand{\eeq}{\end{equation}}
\newcommand{\be}{\begin{eqnarray}}
\newcommand{\ee}{\end{eqnarray}}

\begin{document}

\title{A derivation of the master equation from path entropy maximization}
\author{Julian Lee}
\email{jul@ssu.ac.kr}
\affiliation{Department of Bioinformatics and Life Science, Soongsil University, Seoul, Korea}
\author{Steve Press\'e}
\email{stevenpresse@gmail.com}
\affiliation{Department of Pharmaceutical Chemistry, University of California, San Francisco, CA}

\date{\today}

\begin{abstract}
The master equation and, more generally, Markov processes are routinely used as models for stochastic processes. They are often justified on the basis of randomization and coarse-graining assumptions.
Here instead, we derive $n^{th}$-order Markov processes and the master equation as unique solutions to an inverse problem.
In particular, we find that when the  constraints are not enough to uniquely determine the stochastic model,
the $n^{th}$-order Markov process emerges as the unique maximum entropy solution
to this otherwise under-determined problem. 
This gives a rigorous alternative for justifying such models
while providing a systematic recipe for generalizing widely accepted stochastic models usually assumed to follow from first principles.
\end{abstract}

\maketitle
\vspace{-0.3in}

\section{Introduction}

Markov chains ~\cite{vK,Chung} are often the starting point for modeling 
condensed phase stochastic dynamics in biophysics~\cite{szaboirina, caosilbey, bere, hong1, hong2, brown} and beyond ~\cite{wang}. 
Markov chains are approximations of continuous system dynamics. They are often
justified on the basis of coarse-graining approximations~\cite{GM}.
Coarse-graining reduces classical phase space --with phase points dynamics governed by Liouville's equations-- to a discrete set of states --with stochastic hopping between states determined by stationary transition probabilities. 
Such coarse-graining methods have recently been used
to show how Markov models can describe the continuous dynamics of biomolecules evolving in complex potential landscapes \cite{noe, bow, kasson}.


A very different approach to stochastic dynamics is due to Filyukov and Karpov~\cite{FK} and later Jaynes~\cite{jaynes}. Using this approach, models for a stochastic dynamics can be inferred as unique solutions to an inverse problem. To be clear, by model we mean the probability distribution for the stochastic paths the system can take.

Normally, the number of stochastic paths greatly outnumbers 
the constraints imposed.
To find a unique solution to this under-determined problem we ask:
which model is it that not only satisfies the limited experimental constraints but also
maximizes the entropy for the path probabilities?
As Shore and Johnson~\cite{shore} showed, 
this is exactly equivalent to finding a model for the path probabilities which 
satisfies the experimental constraints while satisfying these logical consistency axioms:
1) when $A$ and $B$ are independent data then the model for $P(A$ and $B)$
must reduce to $P(A)P(B)$ and the model for $P(A$ or $B)$ must reduce to $P(A)+P(B)$; 
2) furthermore, any prediction made from the model must be independent of the coordinate system used in the calculation. 

This method of finding a stochastic model is mathematically similar to the maximum entropy principle for determining equilibrium probability distributions~\cite{jaynes57,GD78,stein92,jaynes03}. 
In earlier work, Ge $et$ $al.$ --which extended the work of Stock $et$ $al.$~\cite{Stock} and 
Ghosh $et$ $al.$~\cite{philken}-- showed that the $1^{st}$ order Markov chain emerges as a natural consequence of path entropy maximization. 
Here we generalize this work in many important ways. 
1) We do not limit ourselves to first order Markov processes; 2) we consider 
under which conditions the master equation emerges as a solution to the procedure of path entropy maximization; 3) we consider how different types of constraints affect the emergent model; 4) we consider very general (non-linear) constraints.

To the best of our knowledge, this is the first time the master equation and, more generally, $n^{th}$-order Markov processes are rigorously shown to 
follow from maximum entropy principles. This provides an alternative justification
for the master equation --the basic tool of stochastic physics and biology-- 
which is distinct from standard chemical or mechanistic justifications provided by
van Kampen ~\cite{vK}, Zwanzig ~\cite{zwan}, Gillespie ~\cite{gillespie} and others. 
The master equation assumes from the onset a dynamics described by stationary
transition probabilities and time-varying state occupation probabilities. 
Here we only assume data of a specific type is available and the basic 
logical consistency axioms required to justify maximum entropy as an inference tool ~\cite{shore}. Posing the master equation as the solution of an inverse problem is significant because possible generalizations to the master equation are now derivable within this
formalism. These generalizations can then be justified on the firm axiomatic basis of
provided by Shore and Johnson.

\section{Markov model of $n^{th}$ order: definitions and notations}

In this section, we briefly introduce the mathematical notation necessary
for the remainder of the paper.
Now, consider a stochastic process in discrete time.
Let the index $i_t$ denote the state of the system at time $t$ along the path $C$  from time $0$ to $T$
where $C=\{i_0,i_1,i_2, \cdots i_T\}$.
The probability distribution of path $C$ is
\be
P(C)=p(i_0,i_1, \cdots i_T) \label{pathprob}
\ee
An $n$-point joint probability is defined as follows
\be
p(a_1, \cdots a_n; t) \equiv \sum_{i_0, i_1, \cdots i_{t-n}, j_1, j_2, \cdots j_{T-t}} p( i_0, i_1, \cdots i_{t-n},  a_1, \cdots a_n , j_1, j_2, \cdots j_{T-t}) \label{joint}
\ee
where, for sake of generality,  $a_1, \cdots a_n$ are neither the first nor the last $t$ indices of 
$p(i_0,i_1, \cdots i_T)$.
The explicit time index is required, as the result depends on which indices are summed over. 
Conditional --also called transition-- probabilities are obtained by invoking Bayes' theorem: 
\be
 p(i_0, \cdots i_{t-1} \to i_t) \equiv p(i_0, \cdots i_{t-1} | i_t) = \frac{p(i_0, \cdots i_t)}{p(i_0, \cdots i_{t-1})} 
\label{condpath}.
\ee 
We call $p(i_0, \cdots i_{t-1} \to i_t)$ a transition probability. 
When the transition probability depends only on the previous $n$-time steps
\be
p(i_0, \cdots i_{t-1} \to i_t) = p(i_{t-n},i_{t-n+1} \cdots i_{t-1} \to i_t; t) \equiv \frac{p(i_{t-n}, \cdots i_t;t)}{p(i_{t-n}, \cdots i_{t-1};t)}, \label{trans}
\ee
the process is called an {\it $n^{th}$-order Markov process}.
When the transition probability is time-independent, it is called a {\it time-homogeneous Markov process}. 
When no specification is given, a Markov process is assumed $1^{st}$-order, time-homogeneous.

\section{Derivation of First order Markov process with linear constraints}

Here we show how the first order Markov process is derived from path entropy maximization.
We begin with the definition of path entropy
\be
H = -\sum_{\{ i_0, i_1, \cdots i_T \}} p(i_0,i_1, \cdots i_T) \log p(i_0,i_1, \cdots i_T). \label{pinfo_disc}
\ee
We consider  $N_1$ ands $N_2$ linear constraints on one and two-point probabilities, respectively:
\be
F^{(\alpha)}_0 & \equiv & \sum_{t=0}^T \sum_{i_t} \varepsilon^{(\alpha)}_{i_t} p(i_t ; t) - (T+1) E^{(\alpha)}_0 =0 \quad (\alpha=1, \cdots N_1) \nonumber\\
F_1^{(\gamma)} &\equiv & \sum_{t=0}^{T-1} \sum_{i_t i_{t+1}} J^{(\gamma)}_{i_t i_{t+1}} p(i_t; t) p(i_t \to i_{t+1} ; t) - T J^{(\gamma)}_0 =0. \quad (\gamma=1, \cdots N_2) \label{linconst}
\ee
and a normalization condition
\be
\sum_{\{ i_0, i_1, \cdots i_T \}} p(i_0,i_1, \cdots i_T)=1 \label{normconst}.
\ee
These constraints are imposed using Lagrange multiplier. That is, the Lagrange multiplier terms are added to the path entropy as follows:
\be
&&-\sum_{\{ i_0, i_1, \cdots i_T \}} p(i_0,i_1, \cdots i_T) \log p(i_0,i_1, \cdots i_T) - \sum_{\alpha=1}^{N_1} \beta_\alpha \left( \sum_{t=0}^T \sum_{i_t} \varepsilon^{(\alpha)}_{i_t} p(i_t ; t) - (T+1) E^{(\alpha)}_0 \right)  \nonumber\\
 && + \sum_{\gamma=1}^{N_2} \nu_\gamma \left( \sum_{t=0}^{T-1} \sum_{i_t i_{t+1}} J^{(\gamma)}_{i_t i_{t+1}} p(i_t, i_{t+1} ; t+1) - T J^{(\gamma)}_0 \right)
+( \rho +1) \left(\sum_{\{ i_0, i_1, \cdots i_T \}} p(i_0,i_1, \cdots i_T)-1\right). \label{target}
\ee
Extremizing Eq.~(\ref{target}) with respect to $p(i_0,i_1, \cdots i_T)$, we obtain 
\be
- \log p(i_0,i_1, \cdots i_T) - \sum_\alpha \beta_\alpha \sum_{t=0}^T \varepsilon^{(\alpha)}_{i_t} + \sum_\gamma \nu_\gamma  \sum_{t=0}^{T-1} J^{(\gamma)}_{i_t i_{t+1}}
+ \rho = 0.  \label{vareq}
\ee
The Lagrange multipliers introduced in Eq.~(\ref{target}) are determined by additional equations
which come from taking the variation of Eq.~(\ref{target}) with respect to these Lagrange multipliers. The solution to Eq.~(\ref{vareq}) is expressed in terms of the Lagrange multipliers as follows
\be
 p(i_0,i_1, \cdots i_T) &=& \exp\left(\rho - \sum_\alpha \beta_\alpha \sum_{t=0}^T \varepsilon^{(\alpha)}_{i_t} + \sum_\gamma \nu_\gamma  \sum_{t=0}^{T-1} J^{(\gamma)}_{i_t i_{t+1} } \right) \nonumber\\
&=&  \exp(\rho) v(i_0) G(i_0, i_1) G(i_1, i_2) \cdots G(i_{T-1}, i_T) v(i_T)
\label{tottraj}
\ee
where the elements of the vector ${\bf v}$, ${v(i)}$, and the elements
of the transfer matrix ${\bf G}$, $G(i, j)$, are defined as follows 
\be
v(i) &=& \exp\left(-\sum_\alpha \beta_\alpha \varepsilon^{(\alpha)}_i/2\right) \nonumber\\
G(i, j) &=&   \exp\left(-\sum_\alpha \beta_\alpha \varepsilon^{(\alpha)}_i/2 + \sum_\gamma \nu_\gamma J^{(\gamma)}_{ij}-\sum_\alpha \beta_\alpha \varepsilon^{(\alpha)}_j/2\right). \label{twopoint}
\ee
The $m$-point joint probability distribution, Eq.~(\ref{joint}), is obtained from
Eq.~(\ref{tottraj}) by summing over indices $ i_{t-m}, i_{t+1}, \cdots i_T$ as follows
\be
p(a_1, \cdots a_m; t) &=&  \sum_{i_0, \cdots i_{t-m}, i_{t+1}, \cdots i_T}  p(i_0,i_1, \cdots i_{t-m},  a_1, \cdots, a_m, i_{t+1}, \cdots,  i_T) \nonumber\\
&=& \exp(\rho) [{\bf v^\dagger} {\bf G}^{t-m+1}](a_1) G(a_1, a_2) G(a_2, a_3) \cdots G(a_{m-1}, a_{m}) [{\bf G}^{T-t} {\bf v} ](a_m) \nonumber\\
&=& \frac{ [{\bf v^\dagger} {\bf G}^{t-m+1}](a_1) G(a_1,  a_2) G(a_2, a_3) \cdots G(a_{m-1}, a_{m}) [{\bf G}^{T-t} {\bf v} ](a_m)}{{\bf v^\dagger} {\bf G}^{T} {\bf v}}. \label{mjoint}
\ee
Therefore combining Eq.~(\ref{trans}) and  Eq.~(\ref{mjoint}), we have
\be
p(a_1, \cdots a_{m} \to a_{m+1}; t) &=& \frac{\exp(\rho) [{\bf v}^\dagger {\bf G}^{t-m}](a_1) G(a_1, a_2) \cdots G(a_{m}, a_{m+1}) [{\bf G}^{T-t} {\bf v}](a_{m+1})}
{\exp(\rho) [{\bf v}^\dagger {\bf G}^{t-m}](a_1) G(a_1,  a_2)  \cdots G(a_{m-1}, a_{m})[ {\bf G}^{T-t+1} {\bf v}] (a_{m}) } \nonumber\\
&=& \frac{G(a_{m}, a_{m+1}) [{\bf G}^{T-t} {\bf v}](a_{m+1})}{ [{\bf G}^{T-t+1} {\bf v}](a_{m})} = p(a_{m} \to a_{m+1}; t). \label{markov_tran}
\ee 
The above is indeed a $1^{st}$ order Markov process though the transition probability has explicit time dependence. 

The $1^{st}$ order Markov property was also derived in Ref.~\cite{HG} for the special case of constraining one-point and two-point statistics which we now define. One-particle statistics, $F_0^{(m)}$ coincide with  
\be
\varepsilon^{(\alpha)}_i = \delta_{i, \alpha} \quad (\alpha = 1, \cdots N)
\ee
where the index $\alpha$ of the constraint now goes over each state of the system, $N$ being their total number of such states. This constraint simply counts the number of times state $\alpha$ is visited over the course of the trajectory (i.e. this constraint is $1$ when state index $i$ is identical to $\alpha$).
Likewise, two-point statistics corresponds to imposing these $F_1^{(\tau,\sigma)}$'s 
\be
J^{(\tau,\sigma)}_{i,j} = \delta_{i,\tau} \delta_{j,\sigma} \quad (\tau,\sigma = 1, \cdots N)
\ee
where we labelled the constraint by double indices $(\tau,\sigma)$ instead of the single index $\gamma$ for notational convenience. This again simply counts the number of transitions from state $\tau$ to $\sigma$ over the course of the trajectory.

\section{Derivation of the time-homogeneous master equation}

Recall that a master equation requires time dependent state probabilities and
time-independent transition probabilities.
Under what conditions are such approximations valid?
To answer this question we apply the Perron-Frobenius theorem
~\cite{Perron, Frob1, Frob2, Frob3, matrix, Markov1, Markov2, Markov3} 
to the ${\bf G}$ transfer matrix of the previous section--a square matrix which by construction 
is of size $N\times N$ and has positive elements.
According to the theorem, ${\bf G}$ satisfies the following properties:

\hspace{-0.15in}(1) It has a positive real eigenvalue $r$, called the Perron-Frobenius eigenvalue, such that any other eigenvalue $\lambda$ is strictly smaller than $r$ in absolute value, $|\lambda| < r$.\newline
(2) There is a left eigenvector ${\bf y^\dagger} = (y_1, \cdots y_N)$ for $r$ with positive components. That is, ${\bf y^\dagger} {\bf G} = r {\bf y^\dagger}$ and $y_i > 0$ for all $i$.  Similarly, there is a right eigenvector ${\bf z}$ with positive components, such that ${\bf G}{\bf z} = r {\bf z}$ and $z_i>0$ for all $i$.\newline
(3) Left and right eigenvectors with eigenvalue $r$ are non-degenerate. \newline
(4) $\lim_{T \to \infty} \frac{{\bf G}^T}{r^T} = {\bf z}{\bf y^\dagger}$

Now re-consider Eq.~(\ref{markov_tran}) where
\be
p(a_{m} \to a_{m+1}; t)=
\frac{G(a_{m}, a_{m+1}) [{\bf G}^{T-t} {\bf v}](a_{m+1})}{ [{\bf G}^{T-t+1} {\bf v}](a_{m})}. 
\label{limvec1}
\ee
Since the vector ${\bf v}$ has only non-negative elements, both  ${\bf G^T}{ \bf v}/r^T$ and  
${ \bf v}^\dagger{\bf G^T}/r^T$ have a well-defined non-zero limit for $T \to \infty$,
\be
\lim_{T \to \infty} \frac{{\bf G}^{T} {\bf v}}{r^T}  =  {\bf z} ({\bf y}^\dagger {\bf v})  ; \hspace{0.15in}
\lim_{T \to \infty} \frac{{\bf v}^\dagger {\bf G}^T}{r^T}  = ({\bf v}^\dagger {\bf z}){\bf y}^\dagger. 
\label{limvec2}
\ee
Therefore, taking the limit $T-t \to \infty$ of Eq.~(\ref{limvec1}) and using Eq.~(\ref{limvec2}), we find
\be
 p(a \to b) = \frac{G(a, b) z(b)}{r z(a)}. \label{stattrans}
\ee
That is, the transition probability is time-independent in this limit.  
However, from Eq.~(\ref{mjoint}), the  $m$-point joint probabilities are still explicitly time-dependent when $T-t$ is large
\be
p(a_1, \cdots, a_m; t) =  \frac{ [{\bf v^\dagger} {\bf G}^{t-m+1}](a_1) G(a_1, a_2) G(a_2, a_3) \cdots G(a_{m-1}, a_{m}) z(a_m)}{r^{t} {\bf v^\dagger}  {\bf z}} \label{mjoint1}
\ee
and, in particular, this is true for the one-point occupation probability
\be
p(a; t) =  \frac{ [{\bf v^\dagger} {\bf G}^{t}](a) z(a)}{r^{t} {\bf v^\dagger}  {\bf z}}. \label{one1}
\ee   

Thus maximizing the path entropy under the linear constraint Eq.~(\ref{linconst}) up to two-point probabilities, which are imposed for infinite duration into the future ($T-t \to \infty$), we obtain a time-homogeneous Markov process which is described by 1) time-independent transition probabilities and 2) time-dependent one-point occupation probabilities.
The resulting evolution equation for this time-homogeneous Markov process 
\be
p(a; t+1) = \sum_b p(b; t) p(b \to a)  \label{master}
\ee 
is the celebrated master equation.

Note the asymmetry in time: the transition probability as well as the joint probabilies are time dependent when the limit of $t \to \infty$ is taken but $T-t$ is kept finite. This is simply due to the fact that the transition probability  $p(b \to a)$ is defined in a time-asymmetric manner.   

The last limit to consider is the stationary case, when both $T-t$ and $t$ are large.  Then the $m$-point joint probability of Eq.~(\ref{mjoint}) reduces to
\be
p(a_1, \cdots a_m) = \frac{y(a_1)  G(a_1, a_2) G(a_2, a_3) \cdots G(a_{m-1}, a_{m}) z(a_m)}{r^{m-1} {\bf y}^\dagger {\bf z}} \label{statm}
\ee
which is independent of time as are the state occupation probability or any conditional probability derived 
from Eq.~(\ref{statm}). This is to
be expected, since we have the time translation invariance in the stationary limit, and the same symmetry
should appear in the probabilities in the absence of additional information.

Stationarity also trivially follows  when the constraints themselves are stationary, which are much stronger conditions than those in Eq.~(\ref{linconst})\footnote{
The stationary process is also obtained when the constraint are imposed at each point in time:
\be
F^{(\alpha)}_0 (t) &=&  \varepsilon^{(\alpha)}_{i_t} p(i_t ; t) -  E^{(\alpha)}_0 =0 \quad (\alpha=1, \cdots N_1) , (t=0, \cdots T)\nonumber\\
F_1^{(\gamma)}(t) &=&  \sum_{i_t i_{t+1}} J^{(\gamma)}_{i_t i_{t+1}} p(i_t; t) p(i_t \to i_{t+1} ; t) - J^{(\gamma)}_0 =0. \quad (\gamma=1, \cdots N_2) (t=0, \cdots T-1)
\ee
Our result shows that the weaker constraint Eq.~(\ref{linconst}) can achieve this so long as $0 \ll t,  T-t$.}.
Eq.~(\ref{statm}) was also derived in the large $T$ limit with ($m=T$) in Ref.~\cite{CecileJSM} though the stationary Markov process was assumed from the onset therein. Likewise, the $1^{st}$ order Markov process was derived  in Ref.~\cite{HG} from path entropy maximization for the special case of pair statistics constraints, but neither conditions for the time-homogeneous process nor stationarity were discussed\footnote{ Adapted to our notation, it is stated underneath of Eq. (11) of Ref.~\cite{HG}, that $p(a,b) \propto G(a,b)$, implying that $p(a,b)$ is time-independent. However, since $p(a,b) = \frac{ [{\bf v^\dagger} {\bf G}^{t-1}](a) G(ab)  [{\bf G}^{T-t} {\bf v} ](b)}{{\bf v^\dagger} {\bf G}^{T} {\bf v}}$ from Eq.(\ref{mjoint}), this is only 
strictly correct when $T-t$ and $t$ are both large.}.

\section{Time-Homogeneous Markov processes  with an arbitrary initial condition}

Often data comes not only in the form of state occupation probabilities
(e.g. how long during the course of a single molecule fluorescence experiment did 
a protein dwell in its compact isoform) or transition probabilities.
Data may also be available in the form of conditions at different points in time
(e.g. the sample is pumped into a photoexcited state at time $t=0$).
Are our conclusions on time-homogeneity from the previous section robust to initial, final or other such conditions? In this section, we briefly show when the time-homogeneity of transition probability depends on such conditions.

Consider an arbitrary condition imposed at time $\tau$
\be
p(a; t=\tau) = \pi(a). \label{initial}
\ee 
We then add the term $\sum_a \lambda(a)  (p(a; \tau) - \pi(a))$ 
with Lagrange multipliers $\lambda(a)\ (a=1, \cdots N)$ to the 
constrained entropy, Eq.~(\ref{target}).
As before, setting the variation with respect to $p(i_0,i_1, \cdots i_T)$ to zero yields
\be
 p(i_0,i_1, \cdots i_T) &=& \exp(\rho+ \lambda(i_\tau) - \beta \sum_{t=0}^T \varepsilon_{i_t} + \nu  \sum_{t=0}^{T-1} J_{i_t i_{t+1} } ) \nonumber\\
&=&  \exp(\rho+\lambda(i_\tau)) v(i_0) G(i_0, i_1) G(i_1, i_2) \cdots G(i_{T-1}, i_T) v(i_T) \nonumber\\
&=&  \frac{ v(i_0) \pi(i_\tau) G(i_0, i_1) G(i_1, i_2) \cdots G(i_{T-1}, i_T) v(i_T)}
{\sum_{j_0 \cdots j_T}  v(j_0) \pi(j_\tau)  G(j_0, j_1) G(j_1, j_2) \cdots G(j_{T-1}, j_T) v(j_T)}
\ee
where in the last line we used the normalization condition Eq.~(\ref{normconst}) 
to eliminate $\rho$ and the  initialization constraint Eq.~(\ref{initial}) to eliminate $\lambda$. 
We now have
\be
&& \tau \le t-m+1:\nonumber\\
&&p(a_1, \cdots a_m; t) 
=   \frac{ \sum_a [{\bf v^\dagger} {\bf G}^{\tau}](a) \pi(a) [{\bf G}^{t-\tau-m+1}](a, a_1) G(a_1, a_2)  \cdots G(a_{m-1}, a_{m}) [{\bf G}^{T-t} {\bf v} ](a_m)}{\sum_b [{\bf v^\dagger} {\bf G}^{\tau}](b)  \pi(b) [{\bf G}^{T-\tau } {\bf v}](b)}   \nonumber\\
&&t-m+1 < \tau \le t:\nonumber\\
&&p(a_1, \cdots a_m; t) \nonumber\\
&&=\frac{  [{\bf v^\dagger} {\bf G}^{t-m+1}](a_1)  G(a_1, a_2)  \cdots G(a_{\tau-t+m-1}, a_{\tau-t+m})\pi(a_{\tau-t+m})  }{\sum_b [{\bf v^\dagger} {\bf G}^{\tau}](b ) \pi(b) [{\bf G}^{T-\tau } {\bf v}](b)} \nonumber\\
&& \times G(a_{\tau-t+m}, a_{\tau-t+m+1})  \cdots  G(a_{m-1}, a_{m}) [{\bf G}^{T-t} {\bf v} ](a_m)  \nonumber\\
&& t < \tau:\nonumber\\
&&p(a_1, \cdots a_m; t) 
= \frac{ \sum_a [{\bf v^\dagger} {\bf G}^{t-m+1}](a_1)   G(a_1, a_2)  \cdots G(a_{m-1}, a_{m})}{\sum_b [{\bf v^\dagger} {\bf G}^{\tau}](b)  \pi(b) [{\bf G}^{T-\tau } {\bf v}](b)} \nonumber\\
&&  \times [{\bf G}^{\tau-t}](a_m, a)\pi(a)  [{\bf G}^{T-\tau} {\bf v} ](a).
\label{mjoint2}
\ee
Using the definition of the transition probability from Eq.~(\ref{trans}) we find
\be
&& \tau< t: \nonumber\\
&&p(a_1, \cdots a_{m} \to a_{m+1}; t) 
= \frac{G(a_{m}, a_{m+1}) [{\bf G}^{T-t} {\bf v}](a_{m+1})}{ [{\bf G}^{T-t+1} {\bf v}](a_{m})}  \nonumber\\
&& \tau \ge t: \nonumber\\
&&p(a_1, \cdots a_{m} \to a_{m+1}; t) 
= \frac{G(a_{m}, a_{m+1}) \sum_a [{\bf G}^{\tau-t}](a_{m+1}, a) \pi(a) [{\bf G}^{T-\tau} {\bf v}](a)}{ \sum_b  [{\bf G}^{\tau-t+1}](a_m, b) \pi(b)  [{\bf G}^{T-\tau} {\bf v}]( b)}.
\label{markov_tran2}
\ee 
We notice that the indices $a_1, \cdots a_{m-1}$ have dropped out from the right hand side
of Eq.~(\ref{markov_tran2}). We can therefore write
\be
 p(a_1, \cdots a_{m} \to a_{m+1}; t) = p(a_{m} \to a_{m+1}; t),
\ee
showing that, once more, we have a $1^{st}$ order Markov process. Furthermore,
the transition probability for $t > \tau$ has exactly the same form as  Eq.~(\ref{markov_tran}), independent of the intial condition ${\bf \pi}$.  It is therefore time-homogeneous under the limit of large $T-t$.
The same is not true of $t\le \tau$, where the transition probability always depends on 
the specified condition and time-homogeneity requires both large $T-\tau$ and $\tau-t$. As noted earlier, this time-asymmetry is a natural consequence of the fact that the definition of the transition probability itself is time-asymmetric.

\section{General Derivation of $n^{th}$-order Markov Process from path entropy maximization}

In this section we generalize the arguments of the previous section 
in two important ways: 1) we consider 
constraints on the data up to $n+1$-point probabilities
\be
F^{(\alpha)}(\{p(i ; t ) \} ,\{  p(i \to j; t) \} , \cdots \{ p(i_0, \cdots i_{n-1}  \to i_n ; t)\}) = 0. \label{multi_const}
\ee
and 2) we do not assume that the constraints $F^{(\alpha)}$ 
are linear functions of their arguments (as was the case for Eq.~(\ref{linconst}) ).

Provided constraints are linear --as was the case in Eq.~(\ref{linconst})-- most of the arguments in the previous sections are generalizable to $n^{th}$-order Markov processes. 
Indeed, the path probability would be described by the multiplication of rank-$(n+1)$ tensors rather than matrices, like Eq.~(\ref{mjoint}). The $n^{th}$-order Markov process would follow immediately though the derivation of the time-homogeneity of various transition probabilities would require the difficult task of applying an analogue of the Perron-Frobenius theorem for general tensors. 

Since we want to derive the $n^{th}$-order Markov process for fully general constraints,
as given by Eq.~(\ref{multi_const}), we take a different route.
We first  express the path probability $p(i_1,i_2, \cdots i_T)$ in terms of the conditional probabilities:
\be
p(i_0,i_1, \cdots i_T) = p(i_0;0)p(i_0 \to i_1;1)p(i_0,i_1\to i_2;2) \cdots p(i_0,i_1 \cdots i_{T-1}\to i_T;T).
\ee
Substituting this expression into Eq.(\ref{pinfo_disc}), we get
\be
H &=&  -\sum_{\{ i_0, i_1, \cdots i_T \}} p(i_0,i_1, \cdots i_T)\left (\log p(i_0;0) + \sum_{t=0}^{T-1} \log p(i_0, \cdots i_t \to i_{t+1};t+1)\right) \nonumber \\
&=& -\sum_{i} p(i;0) \log p(i;0) - \sum_{t=0}^{T-1}  \sum_{\{ i_0, i_1, \cdots i_{t+1} \}} p(i_0,i_1, \cdots i_{t+1};t+1) \log p(i_0, \cdots i_t \to i_{t+1}; t+1) \label{multi_decomp}
\ee
where, in getting from first to second line, we invoked the relation between joint and marginal probabilities; $p(i_0 \cdots i_m;m) = \sum_{i_{m+1} \cdots i_T} p(i_1,i_2, \cdots i_T).$

Now reconsider the constraints given by Eq.~(\ref{multi_const}) imposed from $p(i ; t )$
to $p(i_0, \cdots i_{n-1}  \to i_n ; t)$. We will maximize the entropy, Eq.~(\ref{multi_decomp}),
in two steps: 1) we maximize the entropy with repect to $\{ p(i_0, \cdots i_k ; t) \}\ (k > n)$ , for given values of $\{ p(i_0, \cdots i_k ; t) \}$ with $k \le n$; 2) we then vary the entropy over the remaining variables,  $\{ p(i_0, \cdots i_k ; t) \}\ (k \le n)$. By assumptions, constraints on the data only matter in step 2. 
Furthermore, as we now show, step 1 (the unconstrained maximization) is sufficient to show that
the general path probability reduces to that of an $n^{th}$-order Markov process.

In order to perform step 1, we first invoke the equality
 \be
-\sum_i q_i \log q_i \le -\sum_i q_i \log p_i \label{prineq}
\ee
for arbitrary probability distributions $p_i$ and $q_i$
\footnote{Using the well-known inequality $\log x \le -1+x$ for $x>0$, we see that
 \be
-\sum q_i \log q_i+\sum q_i \log p_i = \sum q_i \log \frac{p_i}{q_i} \le \sum_i q_i (-1+\frac{p_i}{q_i}) = - \sum_i q_i + \sum_i p_i = 0, 
 \ee
proving the inequality Eq.~(\ref{prineq}). This inequality was also invoked in Ref.~\cite{FK}
in a much narrower setting (of deriving a $0^{th}$ order Markov model).}. It follows from Eq.~(\ref{prineq}) that
\be
&&-\sum_j p(i_0, \cdots, i_{m-1} \to j ; t) \log p(i_0, \cdots, i_{m-1} \to j ; t) \nonumber\\
&\le&  -\sum_j p(i_0, \cdots, i_{m-1} \to j ; t) \log p(i_{m-n}, \cdots, i_{m-1} \to j ; t)
\label{inter1}
\ee
Summing both sides of Eq.~(\ref{inter1}) over $i_0, \cdots i_{m-1}$, we find
\be
&&= -\sum_{i_0, \cdots i_{m-1},j} p(i_0, \cdots, i_{m-1}, j ; t) \log p(i_0, \cdots i_{m-1} \to j ; t)\nonumber\\
&\le&  -\sum_{i_0, \cdots i_{m-1},j} p(i_0, \cdots i_{m-1}, j ; t) \log p(i_{m-n}, \cdots i_{m-1} \to j ; t) \label{multi_ineq}
\ee
The above sets a bound on the last term of the path entropy, Eq.~(\ref{multi_decomp}).
Therefore,  for given values of $\{ p(i_0, \cdots i_k ; t) \}$ with $k \le n$, we see that $H$ is maximized for
\be
p(i_0, \cdots, i_{m-1} \to j ; t) = p(i_{m-n}, \cdots, i_{m-1} \to j ; t)\quad (m>n), \label{multi_eq}
\ee
the system now being described by a $n^{th}$-order Markov model where the probability $p(i;t)$ is determined only  by previous $n$ steps of history.

Now Eq.~(\ref{multi_eq}) for the path probability $n^{th}$-order Markov process can be substituted  into the path entropy formula 
Eq.~(\ref{multi_decomp}). Step 2 can be carried forward: the resulting path entropy can be maximized with respect to the remaining variables $p(i_0;t), p(i_0 \to i_1;t), \cdots p(i_0, i_1, \cdots i_{n-1} \to i_n;t)$  under the constraints Eq.~(\ref{multi_const}). 

In summary, we have just shown that $n^{th}$-order Markov processes follow
under very general constraints provided by Eq.~(\ref{multi_const}).
Markov models emerge from the entropy maximization method -- and these
provide immediate and principled generalizations of the ubiquitous master equation.

\section{Discussion}

Markov processes and master equations --the evolution equation describing a $1^{st}$ order time-homogeneous Markov process-- are standard stochastic modeling tools invoked across 
disciplines. Such models are usually justified mechanistically
by coarse-graining arguments or by
assuming quick randomization in space of reactants and products 
(the ``well-stirred" approximation).
Yet it is challenging to ascertain $a$ $priori$ whether any of these conditions actually hold.
Just like maximum entropy has provided an alternative to ergodic theory
for the justification of the equilibrium probability distribution~\cite{jaynes57},
we believe that the path entropy techniques of Filyukov and Karpov~\cite{FK}, and later Jaynes~\cite{jaynes},
provide a compelling axiomatic basis for the Markov process and the master equation.
Here the Markov process emerges as a solution to the following inverse problem:
given measurable $n$-point constraints on a trajectory, what is the least biased 
model for a probability distribution? By least biased, we mean one that, for instance, does not impose correlations in a model when such correlations are not otherwise 
warranted by the data (technically these are the logical consistency axioms of Shore and Johnson). The unique solution to this problem is that
which maximizes the entropy subject to constraints from the data. 

With this formalism, we justify generalizations of the master equation 
on rigorous mathematical grounds.
It is tempting to conjecture whether the $n^{th}$-order Markov process can lead to a time-homogeneous process so long as the constraints are imposed for a time much longer than that of one time step. The proof would require an analogue of the Perron-Frobenius theorem for general tensors, 
an interesting subject for further investigation.


\section{Acknowledgements}
We thank Ken Dill, Kingshuk Ghosh and Hao Ge for useful discussions.


\end{document}